\RequirePackage{rotating}
\documentclass[fleqn, usenatbib]{mnras}

\usepackage[T1]{fontenc}

\DeclareRobustCommand{\VAN}[3]{#2}
\let\VANthebibliography\thebibliography
\def\thebibliography{\DeclareRobustCommand{\VAN}[3]{##3}\VANthebibliography}

\usepackage{graphicx}	% Including figure files
\usepackage{amsmath}	% Advanced maths commands
\usepackage{amssymb}	% Extra maths symbols

\usepackage{flushend}
\usepackage{balance}

\usepackage{float}
\usepackage{caption}
\usepackage{lipsum}

\usepackage{fixltx2e}
\usepackage{xcolor}
\usepackage{tablefootnote}

\usepackage{pdflscape}
\usepackage{longtable}
\usepackage{booktabs}
\usepackage{color}
\usepackage{amssymb}
\usepackage{mathtools}
\usepackage{xspace}
\usepackage{rotating}
\usepackage{appendix}
\usepackage{multirow}
\usepackage{multicol}
\usepackage{array}
\usepackage{subcaption}
\usepackage{comment}
\usepackage{tabularx}
\usepackage{bigstrut}
\usepackage{threeparttable}
\usepackage{afterpage}
\usepackage{capt-of}

\usepackage[normalem]{ulem}
\usepackage{siunitx}

\usepackage{tabularx}
\usepackage{array}
\usepackage{rotating}

\newcommand{\ecyc}[1]{\ensuremath{E_{\rm{C}}}}

\newcommand{\krtext}{\textcolor{black}}
\newcommand{\krrtext}{\textcolor{black}}

\usepackage{newtxtext, newtxmath}

\title[Eclipse flares in LMXBs]{Thermonuclear X-ray bursts across the eclipse transitions in the LMXBs EXO 0748-676 and XTE J1710-281}

\author[Rikame et al.]{
Ketan Rikame$^{1,2}$,\thanks{E-mail: rikame.bhaskar@res.christuniversity.in}
Biswajit Paul$^{2}$,
Rahul Sharma$^{2,3}$, and
V. Jithesh$^{1}$
\\
$^{1}$Department of Physics and Electronics, Christ University, Bangalore 560029. Karnataka, India\\
$^{2}$Raman Research Institute, Astronomy and Astrophysics, C.V. Raman Avenue, Bangalore 560080. Karnataka, India\\
$^{3}$Inter-University Centre for Astronomy and Astrophysics, Post Box No. 4, Ganeshkhind, Pune 411007. Maharashtra, India\\
}

\date{Accepted XXX. Received YYY; in original form ZZZ}

\pubyear{2024}

\begin{document}
\label{firstpage}
\pagerange{\pageref{firstpage}--\pageref{lastpage}}
\maketitle

\begin{abstract}

The primary radiation from thermonuclear X-ray bursts observed in the neutron star low-mass X-ray binary (LMXB) systems can interact with various parts of the binary system. This interaction gives rise to secondary radiation in different wavelength ranges, known as reprocessed emission. In eclipsing LMXBs, the reprocessed emission from the bursts can be examined during eclipses, as the primary emission is blocked and only the reprocessed emission is visible. We searched for bursts during eclipses in the archival \textit{RXTE} data of the eclipsing LMXBs and found them in EXO 0748-676 and XTE J1710-281. In EXO 0748-676, seven bursts were found to occur near eclipse egress, with their tails extending beyond the eclipse, and one such burst was found for XTE J1710-281. We estimate the reprocessing fraction at orbital phases near eclipse egress by modeling the peculiar eclipse bursts detected in both systems, which have tails extending beyond the eclipses. We observe an increasing trend in reprocessing fraction as these eclipse bursts occur closer to the eclipse egress. \krrtext{We discuss the possibilities of reprocessing in the ablated wind from the companion star, the accretion disc, and the disc wind in EXO 0748-676 and XTE J1710-281. Additionally, we observe two decay components in the bursts in EXO 0748-676, which could suggest a complex composition of the accreting fuel. From the burst rise timescales, we place an upper limit on the size of the reprocessing regions in both EXO 0748-676 and XTE J1710-281, finding it comparable to the size of the respective X-ray binaries.}

\end{abstract}


%%%%%%%%%%%%%%%%% BODY OF PAPER %%%%%%%%%%%%%%%%%%

\section{Introduction}

X-ray binary systems consist of a companion star and a compact object in orbit, where matter is transferred from the companion star to the compact object \citep{Lewin1995}. The compact object can be a neutron star or a black hole. X-rays are produced near the compact object as a result of matter being gravitationally accreted from the companion star.  \krrtext{The primary X-rays originate in the innermost regions of the accretion flow, such as the inner disc, the boundary layer (in neutron star systems), or corona. While most of these primary X-rays reach the observer directly, a significant fraction interacts with the surrounding matter - including the accretion disc, the corona, the companion star’s surface, or outflowing material - leading to the production of secondary X-rays.} This process is called X-ray reprocessing, and the secondary X-rays are called reprocessed X-rays. In eclipsing X-ray binary systems, during an eclipse, the companion star blocks the primary X-rays, leaving only the reprocessed emission observable, making these systems ideal for studying reprocessed X-rays \citep{Nafisa2019, Nafisa2024, Rikame2024}.

\krrtext{Low-mass X-ray binaries (LMXBs) are a subclass of X-ray binaries in which the companion star typically has a mass \(\lesssim 1\,M_\odot\) and transfers material to the compact object primarily via Roche-lobe overflow, forming an accretion disc \citep{Lewin1995}.} Thermonuclear (Type-I) X-ray bursts (hereafter referred to as 'bursts') observed in neutron star LMXBs are characterized by a sudden increase in X-ray emission over time scales of a few seconds and an exponential decay over a few tens of seconds \citep{WHGLewin1993, Strohmayer2006, Galloway2008, Galloway2021}. These bursts are caused by the unstable thermonuclear fusion of the infalling material on the surface of the neutron star. The freshly accreted matter is compressed and heated over hours to days to densities and temperatures sufficient for thermonuclear ignition \citep{Kuulkers2003}. The timing characteristics of thermonuclear (Type-I) bursts are generally modeled with a Fast Rise Exponential Decay (FRED) profile. The decay rate of the bursts depends on the thermonuclear fuel. The X-ray spectra during the bursts are consistent with black-body emission from a compact object with a radius of approximately 10 km and a temperature between 2 to 3 keV \citep{Galloway2008, Galloway2021}. The spectral hardening during the rise of the X-ray burst and the subsequent softening during the decay reflects the heating and cooling of the neutron star surface \citep{Kuulkers2003}.

A consequence of the thermonuclear X-ray bursts observed in LMXBs - aside from being directly detected by observers - is the reprocessing of X-ray photons as they interact with various components of the binary system. \krrtext{Bursts can influence the accretion rate in the inner disc, potentially through mechanisms such as Poynting–Robertson drag or irradiation-driven instabilities \citep{Ballantyne2005, Worpel2013, Zhao2022}.} In eclipsing LMXBs, the reprocessed emission from the bursts can be examined during eclipses, as the primary emission is blocked and only the reprocessed emission is visible. \krrtext{Additionally, this reprocessed emission can be studied through simultaneous observations at longer wavelengths, where most of the burst emission is expected to be reprocessed. Such reprocessed burst emission has been reported in the ultraviolet (UV) \citep{hynes2006}, optical \citep{graman2017}, infrared (IR) \citep{Vincentelli2021}, and radio \citep{Russell2024} bands.} In this work, we investigate eclipse X-ray bursts detected in eclipsing LMXBs and estimate the reprocessing fraction and put upper limit on the size of the reprocessing region.

EXO 0748-676 is an eclipsing LMXB discovered with EXOSAT in 1985 \citep{Parmar1985}. It underwent an outburst lasting about 20 years \citep{Parmar1986} and entered quiescence in late 2008 \citep{Degenaar2011}. After 16 years of quiescence, it went into outburst again in July 2024 \citep{Rhodes2024, Bhattacharya2024}. The system has an orbital period of 3.82 hours, with an eclipse duration of approximately 500 seconds \citep{Parmar1986, Wolff2009}. The presence of eclipses and complex dipping activity suggests that the system is viewed at an inclination angle between 75$^\circ$ and 82$^\circ$, which is very close to the accretion disc plane. The mass of the companion is constrained to be approximately 0.5 $M_{\odot}$ \citep{Parmar1986}. \cite{hynes2006} have estimated the projected area of the companion star, assuming a spherical approximation, to be \((2.0 \pm 0.9) \times 10^{21} \, \mathrm{cm}^2\), and the tidal truncation radius to be \((0.50 \pm 0.05) \times 10^{11} \, \mathrm{cm}\). \cite{Wolff2005} estimated the distance to EXO 0748-676 to be 7.7 kpc for a helium-dominated burst photosphere and 5.9 kpc for a hydrogen-dominated burst photosphere. A strong X-ray burst from this source provides evidence of photospheric expansion \citep{Wolff2005}. \cite{Ozel2006} used photospheric radius expansion (PRE) bursts to estimate the mass and radius of the neutron star as $M_{\mathrm{NS}} = 2.10 \pm 0.28 \, M_\odot$ and $R_{\mathrm{NS}} = 13.8 \pm 1.8$ km, respectively. These values were shown to rule out soft equations of state. EXO 0748-676 is a unique LMXB exhibiting all types of variability commonly observed in different LMXBs, such as eclipses, quasi-periodic oscillations (QPOs), and bursts \citep{Bonnet-Bidaud2001}. \cite{Boirin2007} reported the detection of burst triplets and the presence of an additional slow decay component in these bursts using \textit{XMM-Newton} observations of EXO 0748-676.

In addition to X-rays, EXO 0748-676 has been observed in optical and UV wavelengths. \cite{graman2017} examined the optical and X-ray light curves of EXO 0748-676 in its non-bursting state \krrtext{(data excluding thermonuclear burst intervals),} finding no correlations between the two at either reprocessing or orbital timescales. While a significant portion of the optical emission is expected to originate from reprocessing, they proposed that the lack of correlation may be due to structures and structural changes in the accretion disc, which can produce or mask the reprocessed signal to varying degrees. During its quiescent state, \cite{Parikh2021} observed extremely low UV luminosity compared to the X-ray luminosity, with no clear indication that the emissions in the two wavebands are connected. They suggested that the UV and X-ray emissions in quiescence likely have different origins. \cite{hynes2006} observed X-ray bursts in EXO 0748-676 along with strong reprocessed signals in UV and optical wavelengths corresponding to these bursts. The observed lagged and smeared signals in the optical and UV bands are consistent with the standard paradigm of prompt reprocessing, distributed across the accretion disc and the companion star. \cite{bpaul2012} detected multiple simultaneous X-ray and optical bursts in EXO 0748-676, with delayed optical bursts, which aligns with the above interpretation. However, they were unable to find any correlation between burst delay times or burst rise times and the orbital phase, which would be expected if reprocessing occurred primarily on the companion star. \cite{AHKnight2025} reported simultaneous detection of thermonuclear bursts in X-ray and optical wavelengths during the source's recent outbursts in June 2024. They observed that while the rise times in optical and X-ray were similar, the optical decay was faster. Their study suggests that the reprocessing site is likely within a few light-seconds of the X-ray-emitting region, making the companion star, accretion disc, and ablated material as plausible candidates.

XTE J1710-281 is another eclipsing LMXB, discovered with \textit{RXTE} in 1998 \citep{Markwardt1998}. The compact object in this binary system is a neutron star \citep{Markwardt2001}. \cite{Iaria2024} has estimated the mass of the companion star to be 0.22 $M_\odot$. It is located at a distance of approximately 15 kpc and has an inclination of about 80$^\circ$ \citep{Frank1987, Markwardt2001}. The orbital period of XTE J1710-281 is 3.28 hours, with the eclipse phase lasting about 420 seconds \citep{Jain2011}. The source exhibits dipping activity, sharp eclipse transitions \citep{Markwardt1998, Markwardt2001}, and thermonuclear bursts.

Both EXO 0748-676 and XTE J1710-281 exhibit orbital period glitches, a feature not observed in any other LMXB \citep{Wolff2009, Jain2011, Jain2022}.

In this paper, we report the detection of X-ray bursts during the eclipses of EXO 0748-676 and XTE J1710-281 in archival \textit{RXTE} data. We discuss the reprocessing characteristics of these sources by modeling their eclipse burst profiles. \krtext{A recent investigation by \cite{AHKnight2025_02} analysed RXTE and XMM-Newton observations of EXO 0748-676. Their work discussed burst statistics, spectral states, and spectral fits to in-eclipse bursts, with a brief discussion on bursts during eclipse egress transitions. In the present study, the focus is specifically on these eclipse-transition bursts (in EXO 0748-676 and XTE J1710-271), with their eclipse-modified light-curve profiles modeled to quantify the orbital phase-dependent reprocessing fraction. The two studies may therefore be viewed as complementary: \cite{AHKnight2025_02} provide a broad perspective on burst statistics and spectral properties, while the present work offers a targeted examination of the eclipse-transition regime.} The paper is organized as follows: Section \ref{observations_eclipse_burst} outlines the archival data survey and data reduction process. Section \ref{analysis_eclipse_burst} details the eclipse burst identification process (Sections \ref{ecl_burs_id}), the analysis of eclipse burst profile modeling (Sections \ref{3.1}, \ref{fred_sing}, \ref{3.3}, and \ref{fred_doub}) and reprocessing fraction estimation (Sections \ref{ecl_st_est} and \ref{rp_est}). In Section \ref{discussion_eclipse_burst}, we discuss the implications of X-ray burst studies during eclipses, and Section \ref{eclipse_burst_conclusion} provides a summary and conclusion.

\section{Observations and data reduction}
\label{observations_eclipse_burst}

We searched for eclipse bursts across the entire archival \textit{RXTE} dataset, covering the full lifetime of the mission (1996–2011), for the eclipsing LMXBs EXO 0748-676, XTE J1710-281, MXB 1658-298, and GRS 1747-312. The \textit{RXTE}-PCA light curves in the 3–22 keV energy range were extracted using \texttt{ftool-seextrct}. Background counts were estimated with \texttt{ftool-pcabackest}, \krrtext{using the appropriate background model.}

\krrtext{In total, we searched 747 observations with a exposure of 2078 ks for EXO 0748-676, 125 observations with 580 ks exposure for XTE J1710-281, 121 observations with 404 ks exposure for MXB 1658-298, and 227 observations with 830 ks exposure for GRS 1747-312.}

\krrtext{Across these observations, 413, 58, and 24 full eclipses have been reported for EXO 0748-676, XTE J1710-281, and MXB 1658-298, respectively \citep{Wolff2009, Jain2011, Jain2017}. The resulting eclipse exposures are 205 ks for EXO 0748-676, 24 ks for XTE J1710-281, and 22 ks for MXB 1658-298. In GRS 1747-312, the eclipse duration in \textit{RXTE} data is typically about 2.6 ks \citep{Painter2024}. However, most \textit{RXTE} observations of GRS 1747-312 cover only partial eclipses and summing all visible segments yields a total eclipse exposure of approximately 76 ks. A total of (including both in-eclipse and out-of-eclipse bursts) 160 bursts in EXO 0748-676, 46 in XTE J1710-281, 26 in MXB 1658-298, and 7 in GRS 1747-312 are reported in the MINBAR catalogue from the same \textit{RXTE}-PCA data \citep{Galloway2010}.}

\begin{table*}
\centering
\renewcommand{\arraystretch}{1.4} % Adjust the row height
\caption{List of \textit{RXTE} observations of EXO 0748-676 showing eclipse variability}
\label{tab:rxte_exo_0748_676_obslist}
\begin{tabular}{|c|l|c|c|c|c|}
\hline
Sr.No. & OBSID & Date and Time (UT) of Observation Start & Duration (s) & Exposure (s) & \krrtext{Eclipse Variability Characteristics} \\
\hline
01 & 10108-01-06-00 & 1996-08-15 02:04:10 & 2703 & 1655 & Entirely in eclipse \\
02 & 20069-05-05-00 & 1997-06-26 21:26:44 & 2919 & 1617 & Spanning Eclipse+OOE, near eclipse egress \\
03 & 30067-04-03-00 & 1998-08-25 08:58:06 & 2683 & 1975 & Spanning Eclipse+OOE, near eclipse ingress \\
04 & 40039-04-04-00 & 1999-07-05 02:38:07 & 1980 & 1734 & Entirely in eclipse \\
05 & 40039-04-05-00 & 1999-07-05 06:25:07 & 2160 & 1913 & Spanning Eclipse+OOE, near eclipse ingress \\
06 & 40039-06-01-00 & 1999-10-17 03:38:07 & 2580 & 2202 & Spanning Eclipse+OOE, near eclipse egress \\
07 & 50045-01-04-00 & 2000-03-29 06:34:40 & 2819 & 2255 & Spanning Eclipse+OOE, near eclipse egress \\
08 & 50045-03-02-00 & 2000-07-11 11:32:22 & 3600 & 2952 & Spanning Eclipse+OOE, near eclipse ingress \\
09 & 50045-03-05-00G & 2000-07-12 02:52:07 & 4340 & 1629 & Spanning Eclipse+OOE, near eclipse ingress \\
10 & 50045-04-03-00 & 2000-08-29 09:22:36 & 2915 & 2023 & Entirely in eclipse \\
11 & 50045-06-02-00 & 2000-12-17 23:27:30 & 2811 & 1914 & Spanning Eclipse+OOE, near eclipse egress \\
12 & 50045-06-05-00 & 2000-12-18 14:48:07 & 3217 & 1912 & Spanning Eclipse+OOE, near eclipse egress \\
13 & 60047-05-04-00 & 2001-10-25 07:47:06 & 2341 & 1963 & Entirely in eclipse \\
14 & 70048-02-04-00 & 2002-03-18 01:05:04 & 2463 & 1776 & Spanning Eclipse+OOE, near eclipse egress \\
15 & 70048-05-02-00 & 2002-06-29 18:37:39 & 2557 & 1956 & Entirely in eclipse \\
16 & 70048-05-04-00 & 2002-06-30 01:58:44 & 3585 & 2425 & Entirely in eclipse \\
17 & 70048-13-06-00 & 2003-08-17 21:25:50 & 2477 & 1934 & Spanning Eclipse+OOE, near eclipse egress \\
18 & 70048-13-07-00 & 2003-08-18 01:23:07 & 3271 & 1675 & Entirely in eclipse \\
19 & 93074-06-01-00 & 2008-03-19 11:20:06 & 3780 & 2397 & Entirely in eclipse \\
20 & 93074-06-04-00 & 2008-03-19 23:18:11 & 2216 & 1953 & Spanning Eclipse+OOE, near eclipse ingress \\
21 & 92040-04-01-00 & 2008-06-29 07:15:46 & 2763 & 1882 & Entirely in eclipse \\
\hline
\end{tabular}\\
\end{table*}

\begin{table*}
\centering
\renewcommand{\arraystretch}{1.4} % Adjust the row height
\caption{List of \textit{RXTE} observations of XTE J1710-281 showing eclipse variability}
\label{tab:rxte_xte_j1710_281_obslist}
\begin{tabular}{|c|l|c|c|c|c|}
\hline
Sr.No. & OBSID & Date and Time (UT) of Observation Start & Duration (s) & Exposure (s) & \krrtext{Eclipse Variability Characteristics} \\
\hline
01 & 40407-01-03-00 & 1999-03-14 00:35:35 & 4328 & 3520 & Entirely in eclipse \\
02 & 60049-01-03-00 & 2001-11-11 00:07:06 & 5177 & 3591 & Entirely in eclipse \\
03 & 91045-01-01-10 & 2005-11-05 04:06:29 & 2585 & 1918 & Spanning Eclipse+OOE, near eclipse egress \\
\hline
\end{tabular}\\
\end{table*}

\begin{figure*}
\centering
\begin{subfigure}[t]{0.48\textwidth}
    \centering
    \includegraphics[width=\textwidth]{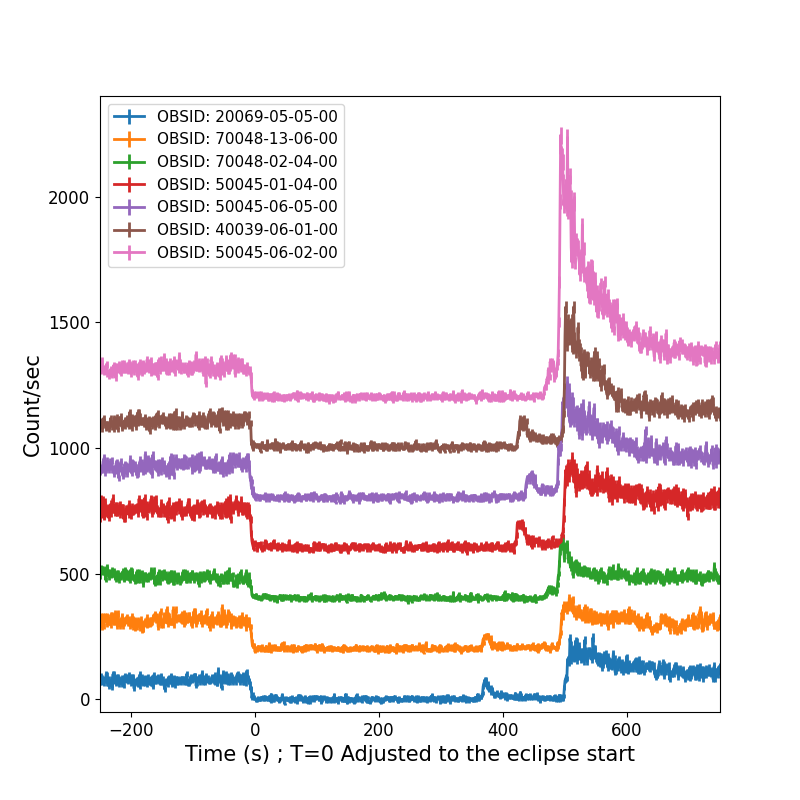}
    %\caption{Distorted eclipse profiles due to bursts occurring during the eclipse transition near the eclipse egress.}
    %\label{fig:exo_0748_676_distorted_eclipse_profile}
\end{subfigure}
\hfill
\begin{subfigure}[t]{0.48\textwidth}
    \centering
    \includegraphics[width=\textwidth]{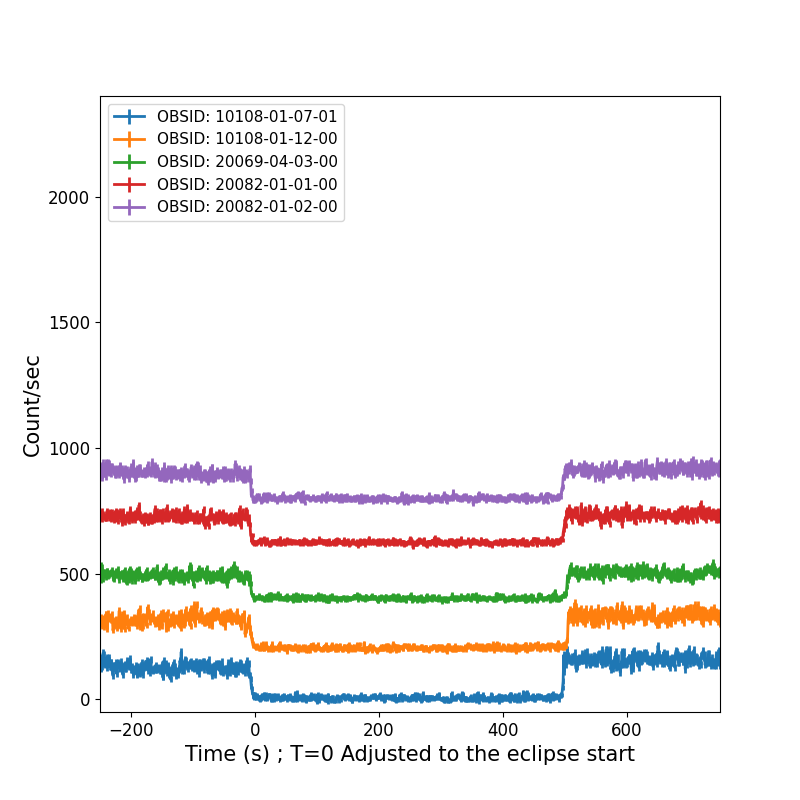}
    %\caption{Representative typical eclipse profiles.}
    %\label{fig:exo_0748_676_normal_eclipse_profile}
\end{subfigure}
\caption{Eclipse profiles in \textit{RXTE} observations of EXO 0748-676. \krrtext{A constant incremental offset of 200 count/s is added to each eclipse profile for better visibility.} \textbf{Left Panel:} Distorted eclipse profiles due to bursts occurring during the eclipse transition near the eclipse egress. \textbf{Right Panel:} Representative typical eclipse profiles.}
\label{fig:exo_0748_676_normal_eclipse_profiles}
\end{figure*}

\section{Analysis}
\label{analysis_eclipse_burst}

\subsection{Eclipse burst identification}
\label{ecl_burs_id}

\krrtext{In the context of burst identification, during eclipses, the primary radiation is obscured by the companion star. As a result, the count rates are significantly lower, and the intrinsic source variability in the light curve becomes comparable to statistical fluctuations in the data. For bursts occurring near the eclipse transitions, the burst shape can be distorted. This makes it challenging to define a universal criterion for identifying such peculiar bursts using automated algorithms. Furthermore, when identifying eclipse durations from orbital ephemerides, EXO 0748-676 and XTE J1710-281 are known to exhibit orbital period glitches \citep{Wolff2009, Jain2011, Jain2022}. The eclipse duration in EXO 0748-676 is also variable \citep{Wolff2009}. In eclipse bursts occurring near the eclipse transition, the eclipse egress profiles are often distorted, making it difficult to determine eclipse durations based solely on count rate variations.}

\krrtext{For these reasons, we have visually examined the light curves of all available \textit{RXTE} observations of EXO 0748-676, XTE J1710-281, MXB 1658-298, and GRS 1747-312 to identify variability during the eclipses.}

We detected 21 eclipse bursts in EXO 0748-676 and 3 eclipse bursts in XTE J1710-281. In MXB 1658-298 and GRS 1747-312, we did not find any eclipse burst. Additionally, EXO 0748-676 went into outburst in June 2024 \citep{Bhattacharya2024}; however, no eclipse bursts were detected in the corresponding \textit{XMM-Newton} and \textit{AstroSat} data \citep{Kashyap2024ATel, Bhattacharya2024ApJ}. Table \ref{tab:rxte_exo_0748_676_obslist} and \ref{tab:rxte_xte_j1710_281_obslist} list the eclipse bursts detected in the \textit{RXTE} observations of EXO 0748-676 and XTE J1710-281, respectively. Among the 21 eclipse bursts detected in EXO 0748-676, 9 bursts occurred entirely within the eclipse duration (see Appendix Figure \ref{fig:exo_0748_676_eclipse_plots}). Five bursts began before the eclipse, with their tails extending into the eclipse duration (see Appendix Figure \ref{fig:exo_0748_676_eclipse_ingress_plots}). The remaining 7 bursts occurred near eclipse egress, with their tails extending beyond the eclipse (see Figure \ref{fig:exo_0478_676_eclipse_burst_fit_02}). The distorted eclipse profiles associated with these bursts are shown in left panel of Figure \ref{fig:exo_0748_676_normal_eclipse_profiles}. In XTE J1710-281, 1 burst with a tail extending beyond the eclipse was observed, while the other 2 bursts occurred entirely within the eclipse duration (see Appendix Figure \ref{fig:xte_j1710_281_eclipse_plots}).
%These bursts, spanning either eclipse ingress or egress, are also reported by \cite{Wolff2009} as unsuitable for eclipse timing analysis due to distorted eclipse profile.

\subsection{Eclipse bursts near the eclipse egress with tails extending beyond the eclipses in EXO 0748-676 and XTE J1710-281}
\label{3.1}

We detected 7 bursts in EXO 0748-676 and 1 burst in XTE J1710-281 where the tail of the burst persisted beyond the eclipse. In principle, such bursts, spanning both the eclipse and out-of-eclipse phases, can be modeled with a FRED profile, applying an appropriate scaling factor to adjust the count rates within the eclipse duration to match the out-of-eclipse emission. This scaling factor can provide insights into the reprocessing efficiency of the system.

We observed 5 bursts in EXO 0748-676 that began before the eclipse, with their tails extending into the eclipse phase. However, modeling the reprocessed burst tails of these bursts is challenging due to poor statistics. For bursts that occur entirely during the eclipse (9 in EXO 0748-676 and 2 in XTE J1710-281), no out-of-eclipse data is available for comparison. Therefore, we have included only the bursts that began a little before the eclipse egress, with tails persisting beyond the eclipse (7 in EXO 0748-676 and 1 in XTE J1710-281), for this analysis. For all bursts, we have uniformly extracted data from 150 seconds before to 450 seconds after the burst start for analysis. The exception is the eclipse burst in EXO 0748-676 corresponding to the \textit{RXTE} OBSID 70048-13-06-00, where the burst decay duration is longer; therefore, a larger duration from the burst decay part was extracted, keeping the total duration the same, i.e., 600 seconds. \krrtext{Additionally, we excluded the data corresponding to the eclipse egress duration from the burst fitting process to ensure that any slope or variability during the egress does not affect the fit to the burst light curves.}

\subsection{Modeling the eclipse burst near the eclipse egress in XTE J1710-281 using a modified FRED profile}
\label{fred_sing}

\krrtext{We modeled the burst profile during eclipse and out-of-eclipse intervals using a modified FRED function. The burst rise is modeled linearly from the start time $t_{\rm start}$ to the peak time $t_{\rm peak}$ with amplitude $A$, followed by an exponential decay with a timescale $\tau$. To account for the eclipse and out-of-eclipse persistent emission, the model includes constant levels $C_{\mathrm{ecl}}$ and $C_{\mathrm{ooe}}$, respectively. The reprocessing fraction ($\eta$) scales the out-of-eclipse burst model relative to the in-eclipse burst model, such that the out-of-eclipse burst model is divided by $\eta$.}

\krrtext{The following modified FRED profile was used for modeling the burst data:}

\krrtext{
\begin{equation*} \label{eq:burst_single_exp}
F_{\mathrm{burst}}(t) =
\begin{cases} 
0 & \text{if } t < t_{\rm start} \\
\left(A \cdot \dfrac{t - t_{\rm start}}{t_{\rm peak} - t_{\rm start}}\right) & \text{if } t_{\rm start} \leq t < t_{\rm peak} \\
\left(A \cdot e^{-\frac{t - t_{\rm peak}}{\tau}}\right) & \text{if } t \geq t_{\rm peak}\\
\end{cases}
\end{equation*}
\vspace{-3em}
\begin{equation}
F_{\mathrm{total}}(t) =
\begin{cases}
F_{\mathrm{burst}}(t) + C_{\mathrm{ecl}} & \text{if } t \text{ is in eclipse duration}, \\[4pt]
\dfrac{F_{\mathrm{burst}}(t)}{\eta} + C_{\mathrm{ooe}} & \text{if } t \text{ is in out-of-eclipse duration}.
\end{cases}
\end{equation}
}

\begin{figure*}
  \centering
  \includegraphics[width=16cm]{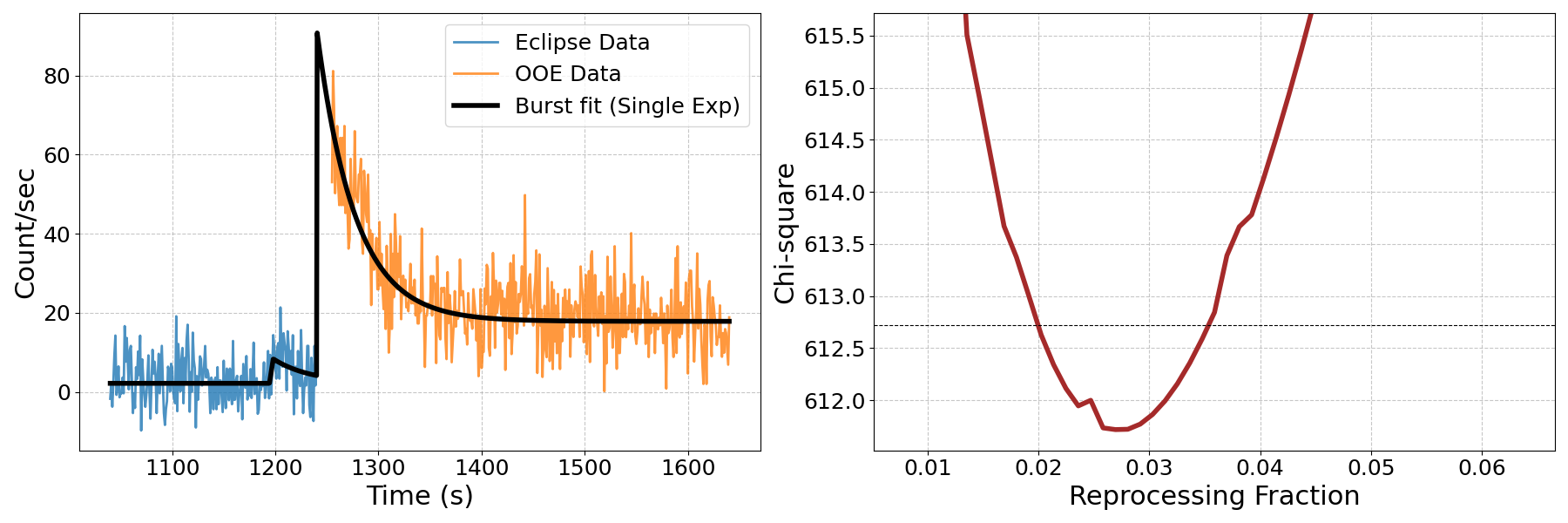}
  \caption{Modeling the eclipse burst in XTE J1710-281 (\textit{RXTE} OBSID: 91045-01-01-10) using a modified profile with a single FRED component. \textbf{Left Panel:} Burst data along with the fitted burst profile. \textbf{Right Panel:} Variation of $\chi^{2}$ with the \krrtext{reprocessing fraction, with the 1$\sigma$ confidence level indicated by a horizontal dashed line.}}
  \label{fig:sing_exp_91045_01_01_10}
\end{figure*}

\krrtext{
\noindent
\begin{minipage}{\linewidth}
\textbf{Where:}
\begin{align*}
& t \text{ is the time.} \\
& t_{\rm start} \text{ is the burst start time.} \\
& t_{\rm peak} \text{ is the burst peak time.} \\
& \tau_{\rm 1} \text{ is the exponential decay timescale of the burst.} \\
& A \text{ is the normalization parameter (amplitude) of the burst.} \\
& C_{\mathrm{ecl}} \text{ and } C_{\mathrm{ooe}} \text{ are the eclipse and out-of-eclipse count rates} \\
& \text{respectively.}\\
& \eta \text{ is the reprocessing fraction, i.e., the fraction of the burst flux}\\
& \text{ observed during eclipse. Out-of-eclipse burst model is divided} \\
& \text{ by } \eta.\\
\end{align*}
\end{minipage}
}
The eclipse burst observed in XTE J1710-281 (\textit{RXTE} OBSID: 91045-01-01-10) was fitted well with the modified FRED profile described above. Figure \ref{fig:sing_exp_91045_01_01_10} shows the fitted eclipse burst profile and the $\chi^2$ variation with repossessing fraction plot, which shows that the reprocessing fraction is well constrained. \krrtext{Table~\ref{tab:xte_j1710_281_fit_para} lists the eclipse burst fit parameters and the fit statistics for the duration near the burst (reduced $\chi^2$), showing that the fit parameters are also well constrained.}

\subsection{Presence of two decay timescales in the bursts of EXO 0748-676}
\label{3.3}

The eclipse bursts observed in EXO 0748-676 could not be fitted well with the modified single exponential decay profile described above.
One possible reason for this poor fit could be that the out-of-eclipse profile near the eclipse egress is not flat and has a characteristic shape that was not considered during the burst profile modeling. To investigate this, several out-of-eclipse profiles of EXO 0748-676 were analyzed, which did not show bursts at the egress, thus providing an indication of what a typical eclipse egress profile looks like. These profiles showed that, unlike the scenario in which a burst occurs during the eclipse transition near the eclipse egress (see left panel of Figure \ref{fig:exo_0748_676_normal_eclipse_profiles}), the typical out-of-eclipse profile near the egress remains flat under normal conditions (see right panel of Figure \ref{fig:exo_0748_676_normal_eclipse_profiles}). Therefore, poor fit of the eclipse bursts is unlikely to be caused by the eclipse egress profile itself.

\begin{figure}
\centering
\includegraphics[width=8cm]{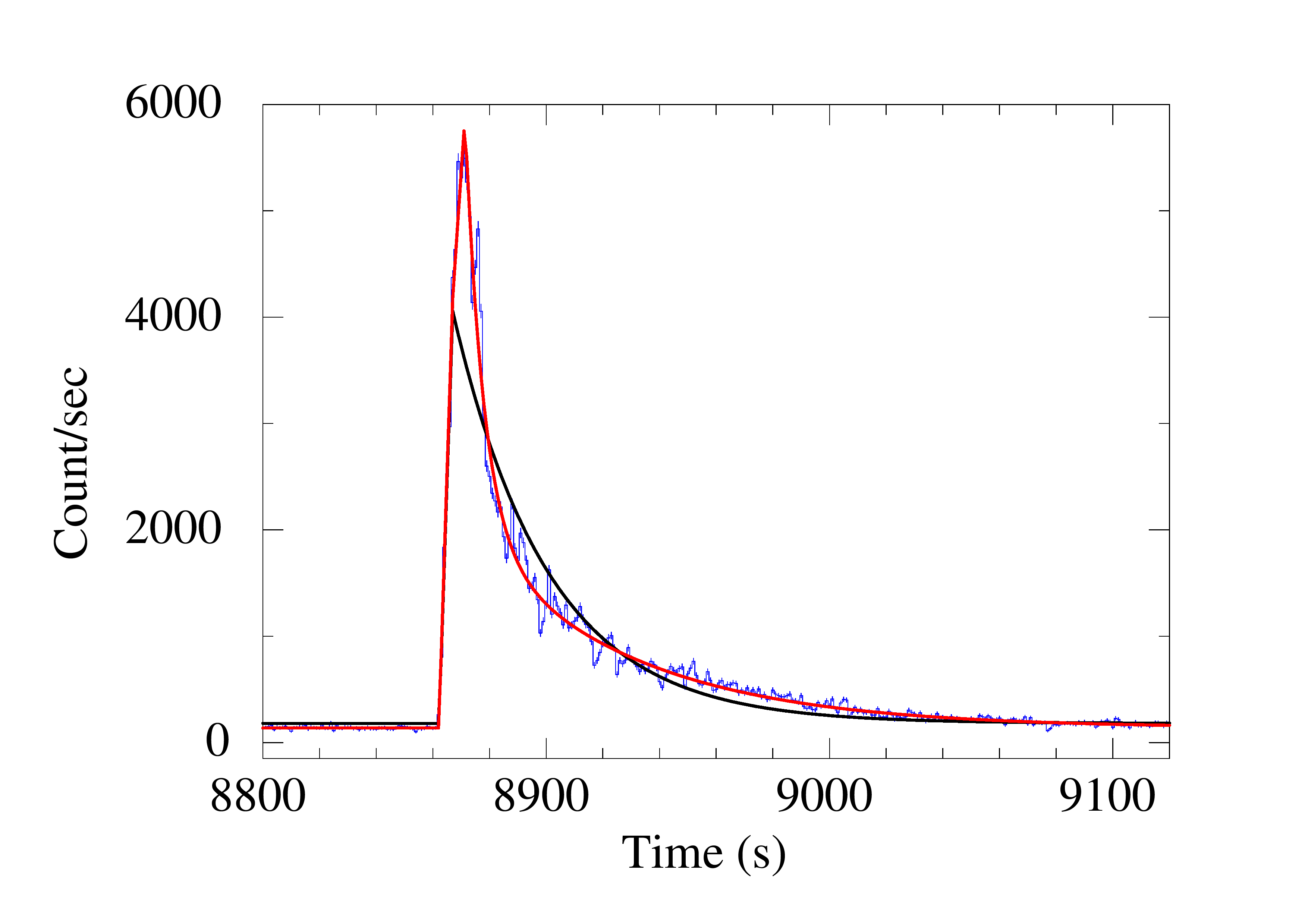}
\caption{Modeling the out-of-eclipse burst in EXO 0748-676 (\textit{RXTE} OBSID: 20082-01-02-000) using a single FRED profile (shown in black) and a double FRED profile (shown in red).}
\label{fig:exo_0748_676_ooe_burst_fit_both}
\end{figure}

\cite{Boirin2007} reported the presence of an additional slow decay component in burst triplets detected in \textit{XMM-Newton} observations of EXO 0748-676. Analysis of several out-of-eclipse bursts in \textit{RXTE} data of EXO 0748-676 also showed a better fit with two FRED components. Figure \ref{fig:exo_0748_676_ooe_burst_fit_both} shows one of the bursts in EXO 0748-676 (OBSID: 20082-01-02-000) fitted with single and double FRED profiles, respectively. \krrtext{For the single FRED fit, the reduced $\chi^{2}$ is $33.43$ for $316$ degrees of freedom. When a second FRED component is included, the reduced $\chi^{2}$ decreases to $7.67$ for $312$ degrees of freedom, indicating a substantial improvement in the fit quality.} Thus, we attempted to model the eclipse bursts in EXO 0748-676 with a modified burst profile that includes two FRED components.

\subsection{Modeling the eclipse burst near the eclipse egress in EXO 0748-676 using a modified FRED profile with two components.}
\label{fred_doub}

\krrtext{We extended the single-component FRED model to include two exponential decay components, allowing us to capture the complex burst profile.}

\krrtext{In this model, the burst rises linearly from $t_{\rm start}$ to $t_{\rm peak}$ with a combined amplitude of $A_{1} + A_{2}$. After the peak, the decay is described by the sum of two exponentials with amplitudes $A_{1}$ and $A_{2}$, and decay timescales $\tau_{1}$ and $\tau_{2}$, respectively. As in the single FRED component model, constant levels $C_{\mathrm{ecl}}$ and $C_{\mathrm{ooe}}$ represent the persistent emission during eclipse and out-of-eclipse intervals, and the reprocessing fraction ($\eta$) scales the out-of-eclipse burst model relative to the in-eclipse burst model by a factor of $1/\eta$.}

\krrtext{
\begin{equation*}
F_{\mathrm{burst}}(t) =
\begin{cases} 
0 & \text{if } t < t_{\rm start} \\\\
\left(A_1 + A_2\right) \cdot \left(\frac{t - t_{\rm start}}{t_{\rm peak} - t_{\rm start}}\right) & \text{if } t_{\rm start} \leq t < t_{\rm peak} \\\\
\left(A_1 \cdot e^{-\frac{t - t_{\rm peak}}{\tau_1}}\right) + \left(A_2 \cdot e^{-\frac{t - t_{\rm peak}}{\tau_2}}\right) & \text{if } t \geq t_{\rm peak} \\\\
\end{cases}
\end{equation*}
%\vspace{-3em}
\begin{equation} \label{eq:burst_double_exp}
F_{\mathrm{total}}(t) =
\begin{cases}
F_{\mathrm{burst}}(t) + C_{\mathrm{ecl}} & \text{if } t \text{ is in eclipse duration}, \\[4pt]
\dfrac{F_{\mathrm{burst}}(t)}{\eta} + C_{\mathrm{ooe}} & \text{if } t \text{ is in out-of-eclipse duration}.
\end{cases}
\end{equation}
}

\krrtext{
\noindent
\begin{minipage}{\linewidth}
\textbf{Where:}
\begin{align*}
& t \text{ is the time.} \\
& t_{\rm start} \text{ is the burst start time.} \\
& t_{\rm peak} \text{ is the burst peak time.} \\
& \tau_{\rm 1} \text{ and } \tau_{\rm 2} \text{ are the exponential decay timescales of the two} \\
& \text{components and } A_1 \text{ and } A_2 \text{ are normalization parameters} \\
& \text{(amplitudes) of the two components.} \\
& C_{\mathrm{ecl}} \text{ and } C_{\mathrm{ooe}} \text{ are the eclipse and out-of-eclipse count rates} \\
& \text{respectively.}\\
& \eta \text{ is the reprocessing fraction, i.e., the fraction of the burst flux} \\
& \text{observed during eclipse. Out-of-eclipse burst model is divided} \\
& \text{by } \eta. \\
\end{align*}
\end{minipage}
}

\begin{figure*} % This will span both columns
    \centering
    \includegraphics[width=\textwidth]{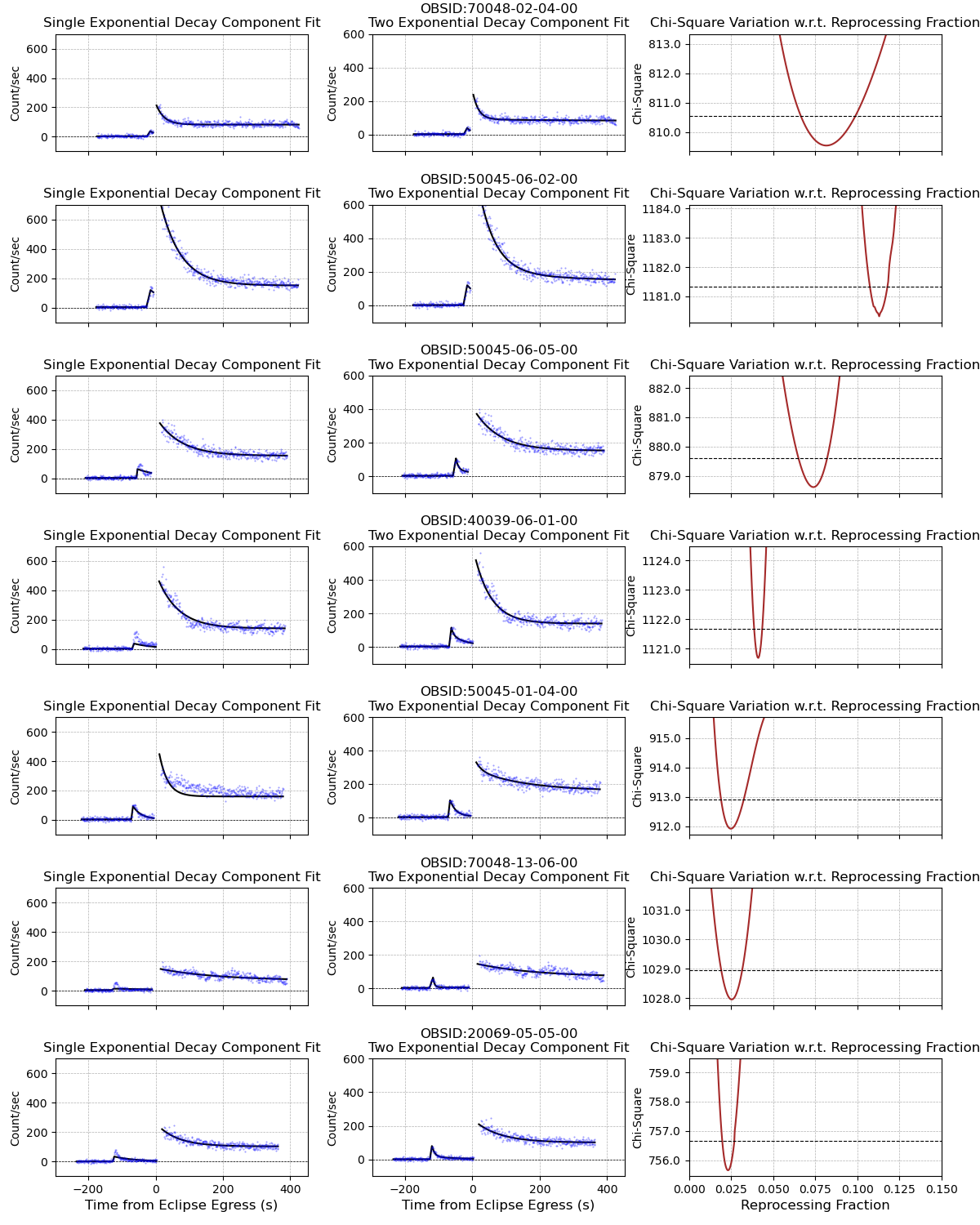}
    \caption[Modeling the eclipse bursts in \textit{RXTE} data of EXO 0748-676, which exhibit tails extending beyond the eclipses, using a modified FRED profile.]{\krrtext{Modeling the eclipse bursts in \textit{RXTE} data of EXO 0748-676, which exhibit tails extending beyond the eclipses, using a modified FRED profile. \textbf{Left panel:} Burst profiles fitted with a single FRED component. \textbf{Middle panel:} Burst profiles fitted with two FRED components. \textbf{Right panel:} Variation of $\chi^{2}$ with the reprocessing fraction, with the 1$\sigma$ confidence level indicated by a horizontal dashed line.}
}
    \label{fig:exo_0478_676_eclipse_burst_fit_02}
\end{figure*}

\krrtext{All the eclipse bursts in EXO 0748-676 with tails extending beyond the eclipse egress showed an improved fit after introducing an additional FRED component, as given in Equation~\ref{eq:burst_double_exp}. Figure~\ref{fig:exo_0478_676_eclipse_burst_fit_02} shows the best-fit burst profiles for these eclipse bursts, fitted with both single (left panel) and double (middle panel) modified FRED components. The variation of $\chi^{2}$ with the reprocessing fraction for the double FRED fit is also shown (right panel), indicating that the reprocessing fraction is well constrained. Table~\ref{tab:exo_0748_676_fit_para} lists the eclipse burst fit parameters and the fit statistics for the duration near the burst (reduced $\chi^2$), obtained from the best fits using both models. An improvement in the reduced $\chi^{2}$ is achieved with the double FRED component model, and the fit parameters are well constrained.}

\subsection{Estimation of the time of the eclipse ingress and egress}
\label{ecl_st_est}

The eclipse ingress profile was fitted with a linear function to estimate the ingress time. The eclipse egress time for both sources was estimated by adding the eclipse duration to the ingress time. For XTE J1710-281, the eclipse duration was taken as 420 seconds, as given by \cite{Jain2011}. Similarly, for EXO 0748-676, the average eclipse duration of \(497.5 \pm 6.0\) seconds, as determined by \cite{Wolff2009}, was used.\\

Table \ref{tab:xte_j1710_281_fit_para} and \ref{tab:exo_0748_676_fit_para} list the eclipse burst fit parameters and the estimates of the time from the eclipse egress to the start of the eclipse bursts for XTE J1710-281 and EXO 0748-676, respectively.\\

\subsection{Reprocessing Fraction as a Function of Time from Eclipse Egress to Burst Start in EXO 0748-676}
\label{rp_est}

\begin{figure} % This will span both columns
    \centering
    \includegraphics[width=8cm]{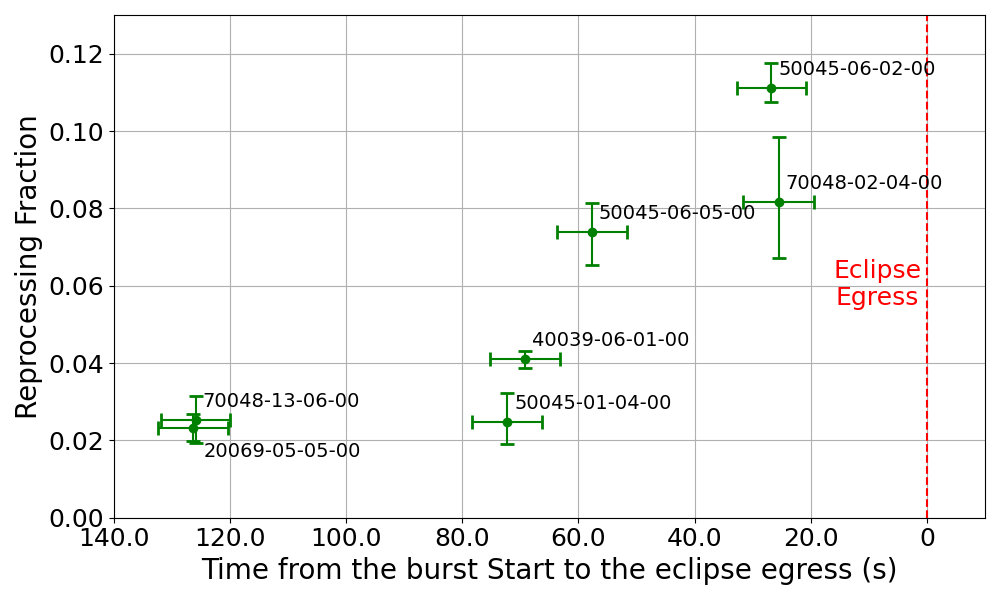}
    \caption{Reprocessing fraction measured in eclipse bursts observed in \textit{RXTE} data of EXO 0748-676, plotted as a function of the time from the burst start to the eclipse ingress.}
    \label{fig:reprocessing_eff_02}
\end{figure}

\krrtext{The reprocessing fraction ($\eta$) is defined as the ratio of the reprocessed burst flux detected during the eclipse to the intrinsic burst flux observed out of eclipse. In our burst modeling, the out-of-eclipse burst model is divided by a factor of $\eta$.} A reprocessing fraction of zero indicates no detection of the reprocessed burst during the eclipse, while a reprocessing fraction of one represents perfect reprocessing, where the reprocessed burst during the eclipse matches the scale of the out-of-eclipse burst. Figure \ref{fig:reprocessing_eff_02} shows the increasing trend in reprocessing fraction in EXO 0748-676 as bursts occur closer to the eclipse egress.

\begin{sidewaystable*}
\centering
%\caption{Two tables stacked on a rotated page.}
\renewcommand{\arraystretch}{1.3}

% First table
\caption{Best-fit parameters for eclipse bursts detected in \textit{RXTE} data of XTE J1710-281, which exhibit tails extending beyond the eclipses. \krrtext{The time values are shifted such that eclipse ingress is assumed to occur at $t=0$.}}
\label{tab:xte_j1710_281_fit_para}
\centering
\fontsize{8}{10}
\renewcommand{\arraystretch}{1.6}
\begin{tabularx}{\linewidth}{
>{\centering\arraybackslash}p{2.2cm}
>{\centering\arraybackslash}p{1.6cm}
>{\centering\arraybackslash}p{1.6cm}
>{\centering\arraybackslash}p{2.4cm}
>{\centering\arraybackslash}p{2.4cm}
>{\centering\arraybackslash}p{1.2cm}
>{\centering\arraybackslash}p{1.2cm}
>{\centering\arraybackslash}p{1.8cm}
>{\centering\arraybackslash}p{1.6cm}
>{\centering\arraybackslash}p{1.8cm}
}
\toprule
OBSID & $t_\mathrm{start}$ & $t_\mathrm{peak}$ & Decay Rate ($\tau$) & Norm ($A$) & $C_{\mathrm{ecl}}$ & $C_{\mathrm{ooe}}$ & Time from burst start to eclipse egress & Reprocessing fraction & $\chi^2_\nu$/dof (Single Decay Component Fit) \\
 & (s) & (s) & (s) & (Count/s) & (Count/s) & (Count/s) & (s) &  &  \\
\midrule
91045-01-01-10 & 378.82 $\pm$ 4.41 & 382.51 $\pm$ 3.30 & 36.76 $\pm$ 3.30 & 6.18 $\pm$ 1.76 & 2.16 $\pm$ 0.49 & 17.83 $\pm$ 0.49 & 41.18 $\pm$ 4.84 & $0.027_{-0.006}^{+0.008}$ & 1.02/178 \\
\bottomrule
\end{tabularx}

\vspace{1cm} % space between the tables

% Second table
\caption{Best-fit parameters for eclipse bursts detected in \textit{RXTE} data of EXO 0748-676, which exhibit tails extending beyond the eclipses. \krrtext{The time values are shifted such that eclipse ingress is assumed to occur at $t=0$.}}
\label{tab:exo_0748_676_fit_para}
\centering
\fontsize{8}{10}
\renewcommand{\arraystretch}{1.6}
\begin{tabularx}{\linewidth}{
>{\centering\arraybackslash}p{2.2cm}
>{\centering\arraybackslash}p{1.2cm}
>{\centering\arraybackslash}p{1.2cm}
>{\centering\arraybackslash}p{1.2cm}
>{\centering\arraybackslash}p{1.2cm}
>{\centering\arraybackslash}p{1.2cm}
>{\centering\arraybackslash}p{1.2cm}
>{\centering\arraybackslash}p{1.2cm}
>{\centering\arraybackslash}p{1.2cm}
>{\centering\arraybackslash}p{1.8cm}
>{\centering\arraybackslash}p{1.6cm}
>{\centering\arraybackslash}p{1.8cm}
>{\centering\arraybackslash}p{1.6cm}
}
\toprule
OBSID & $t_\mathrm{start}$ & $t_\mathrm{peak}$ & Decay Rate 1 ($\tau_1$) & Norm 1 ($A_1$) & Decay Rate 2 ($\tau_2$) & Norm 2 ($A_2$) & $C_{\mathrm{ecl}}$ & $C_{\mathrm{ooe}}$ & Time from burst start to eclipse egress & Reprocessing fraction & $\chi^2_\nu$/dof (Single Decay Component Fit) & $\chi^2_\nu$/dof \newline (Two Decay Components Fit) \\
 & (s) & (s) & (s) & (Count/s) & (s) & (Count/s) & (Count/s) & (Count/s) & (s) &  &  & \\
\midrule
70048-02-04-00 & 471.94 $\pm$ 1.30 & 481.71 $\pm$ 1.28 & 17.03 $\pm$ 2.60 & 35.56 $\pm$ 3.54 & 257.12 $\pm$ 752.11 & 0.77 $\pm$ 0.42 & 3.11 $\pm$ 0.59 & 82.64 $\pm$ 10.43 & 25.06 $\pm$ 6.17 & $0.082_{-0.014}^{+0.017}$ & 1.56/178 & 1.49/176\\
50045-06-02-00 & 470.73 $\pm$ 0.71 & 481.99 $\pm$ 0.92 & 49.09 $\pm$ 7.05 & 110.57 $\pm$ 11.58 & 215.85 $\pm$ 528.75 & 7.64 $\pm$ 11.09 & 2.37 $\pm$ 0.90 & 146.06 $\pm$ 35.20 & 26.27 $\pm$ 6.06 & $0.111_{-0.004}^{+0.006}$ & 3.73/173 & 3.69/171\\
50045-06-05-00 & 439.79 $\pm$ 0.61 & 448.44 $\pm$ 0.72 & 7.11 $\pm$ 2.59 & 67.40 $\pm$ 8.49 & 74.93 $\pm$ 3.81 & 36.91 $\pm$ 6.35 & 2.89 $\pm$ 0.82 & 152.22 $\pm$ 1.95 & 57.21 $\pm$ 6.05 & $0.074_{-0.009}^{+0.007}$ & 2.63/162 & 2.18/160\\
40039-06-01-00 & 428.22 $\pm$ 0.53 & 434.77 $\pm$ 0.74 & 6.84 $\pm$ 3.63 & 45.32 $\pm$ 10.91 & 49.90 $\pm$ 1.72 & 67.20 $\pm$ 6.00 & 3.40 $\pm$ 0.87 & 139.88 $\pm$ 1.54 & 68.78 $\pm$ 6.04 & $0.041_{-0.002}^{+0.002}$ & 5.73/179 & 2.61/177\\
50045-01-04-00 & 425.25 $\pm$ 0.53 & 431.05 $\pm$ 0.64 & 17.72 $\pm$ 1.39 & 96.20 $\pm$ 6.68 & 148.84 $\pm$ 29.76 & 5.15 $\pm$ 2.12 & 4.60 $\pm$ 0.85 & 159.68 $\pm$ 7.13 & 71.75 $\pm$ 6.03 & $0.025_{-0.006}^{+0.007}$ & 7.71/172 & 1.82/170\\
70048-13-06-00 & 371.56 $\pm$ 0.94 & 381.84 $\pm$ 0.85 & 4.36 $\pm$ 1.33 & 58.31 $\pm$ 6.73 & 178.08 $\pm$ 33.12 & 4.11 $\pm$ 1.89 & 4.20 $\pm$ 0.93 & 69.51 $\pm$ 5.55 & 125.44 $\pm$ 6.09 & $0.025_{-0.006}^{+0.006}$ & 2.25/163 & 1.19/161\\
20069-05-05-00 & 371.08 $\pm$ 0.50 & 377.11 $\pm$ 0.58 & 8.13 $\pm$ 1.50 & 65.14 $\pm$ 6.18 & 75.84 $\pm$ 5.88 & 15.84 $\pm$ 2.79 & 0.99 $\pm$ 0.67 & 100.20 $\pm$ 1.64 & 125.92 $\pm$ 6.04 & $0.023_{-0.003}^{+0.004}$ & 2.06/172 & 1.26/170\\
\bottomrule
\end{tabularx}

\end{sidewaystable*}

\section{\krrtext{Discussion}}
\label{discussion_eclipse_burst}

\krrtext{In our search for eclipse bursts in all available archival \textit{RXTE} data of EXO 0748-676, XTE J1710-281, MXB 1658-298, and GRS 1747-312, we found 21 eclipse bursts in EXO 0748-676, 3 eclipse bursts in XTE J1710-281, and no eclipse bursts in MXB 1658-298 or GRS 1747-312.  
Assuming uniform burst occurrence across orbital phases (i.e., no orbital dependence), we can estimate the expected number of eclipse bursts by multiplying the eclipse exposure fraction by the total number of reported bursts for each source. For EXO 0748-676, the total exposure is 2078 ks, with $\sim$205 ks of eclipse exposure ($\sim$10\%) \citep{Wolff2009}. Given 160 reported bursts \citep{Galloway2010}, $\sim$16 bursts are expected during eclipses. We indeed detect 16 bursts that start during eclipses. The remaining 5 eclipse bursts reported in this work have onsets before the eclipse ingress, bringing the total to 21 eclipse bursts. Similarly, for XTE J1710-281, the expected number is $\sim$2, and we observe 3. In MXB 1658-298 and GRS 1747-312, we expect $\sim$1 eclipse burst each, but we detect none. The number of eclipse bursts found does not deviate significantly from what is expected under the consideration of uniform burst occurrence across orbital phases.}

\krrtext{The accretion disc, the surface of the companion star, the ablated wind from the companion star, the disc wind, etc., are possible reprocessing agents in LMXBs such as EXO 0748-676 and XTE J1710-281 \citep{Nafisa2024, AHKnight2025_02}. During the eclipse, the primary emission from the neutron star is blocked by the companion star, and only the reprocessed emission from the visible part of the reprocessing region can be observed. We detected seven bursts at different orbital phases, little before the eclipse egress, in the \textit{RXTE} data of EXO 0748-676, where the tail of the bursts persisted beyond the eclipse. We have shown that, bursts spanning both the eclipse and out-of-eclipse phases can be modeled with a FRED profile, applying an appropriate scaling to adjust the burst count rates within and outside the eclipse duration. In our burst modeling, we define the reprocessing fraction ($\eta$) as the ratio of the reprocessed burst flux detected during the eclipse to the intrinsic burst flux observed out of eclipse. This parameter scales the out-of-eclipse burst model relative to the in-eclipse burst model by a factor of $1/\eta$. The reprocessing fraction thus provides insights into the reprocessing efficiency of the system.}

To investigate the behavior of the reprocessing fraction at various orbital phases just before the eclipse egress, we fitted the eclipse bursts near egress in EXO 0748-676 and XTE J1710-281 with an appropriate burst profile. Thermonuclear bursts are typically modeled using a FRED profile; however, it was found that for the eclipse bursts in EXO 0748-676, two exponential decay components are necessary for an adequate fit. The presence of an additional slow decay component has been observed in the primary bursts of burst triplets during \textit{XMM-Newton} observations of EXO 0748-676 \citep{Boirin2007}. Furthermore, an analysis of several out-of-eclipse bursts in \textit{RXTE} data of EXO 0748-676 indicated a better fit with two exponential decay components. The decay timescale of bursts is influenced by the composition of the thermonuclear fuel being burned, along with factors such as the accretion rate, ignition depth, etc. \citep{Galloway2021}. \krrtext{The presence of two decay components in EXO 0748-676 may suggest a complex composition of the accreting fuel. The single eclipse transition burst in XTE J1710–281, however, was well fitted with a single FRED component.}

Estimating the eclipse egress time is necessary to investigate the variation of the reprocessing fraction at various orbital phases, as the time between the burst start and the eclipse egress provides insight into the orbital position of the neutron star at the time of the burst. Since, for the bursts studied in this work, the eclipse egress profile is also influenced by the burst, modifying the normal egress profile, direct estimation of the eclipse egress time from the profile is difficult. Therefore, we instead estimated the eclipse ingress time by fitting the eclipse ingress profile with a linear function and then estimated the eclipse egress time by adding the eclipse duration to the estimated eclipse ingress time (see Section \ref{ecl_st_est}). \cite{Wolff2009} observed variations in the eclipse duration using \textit{RXTE} data from 1996 to 2001. These variations are not gradual, and adjacent orbits also show large fluctuations in eclipse duration. Thus we used the average value of \(497.5 \pm 6.0\) seconds, as estimated by \cite{Wolff2009}, as the eclipse duration for all eclipse bursts.

\krrtext{Figure~\ref{fig:exo_0478_676_eclipse_burst_fit_02} shows the best-fit burst profiles for bursts in EXO 0748–676 with tails extending into the eclipse egress, fitted with both single (left panel) and double (middle panel) modified FRED components. The variation of $\chi^{2}$ with the reprocessing fraction for the double FRED fit is also shown (right panel). Figure~\ref{fig:sing_exp_91045_01_01_10} shows the fitted eclipse burst profile with a single FRED component (left panel) and the $\chi^2$ variation with reprocessing fraction (right panel) for the only eclipse burst in XTE J1710–281 with a tail extending into the eclipse egress.  Tables~\ref{tab:xte_j1710_281_fit_para} and \ref{tab:exo_0748_676_fit_para} list the eclipse burst fit parameters, the estimates of the time from the eclipse egress to the start of the eclipse bursts, and the fit statistics (reduced $\chi^2$) for XTE J1710–281 and EXO 0748–676, respectively. The reprocessing fraction and all other fit parameters are well constrained in all observations.}

\krtext{The measured reprocessing fractions in EXO~0748-676 range from $2.3\%$ to $11.1\%$, with the errors of the order of 0.5-1\%. \cite{AHKnight2025_02} inferred an average reprocessing fraction of $\sim 2.4\%$, but with a substantially wider range, spanning nearly an order of magnitude in the intrinsic peak burst flux. A likely reason for this difference lies in the respective analysis approaches. In \cite{AHKnight2025_02}, only the out-of-eclipse portions of the egress-split bursts were fitted using an exponential profile. In the present study, the full eclipse-transition bursts were modeled by simultaneously fitting the in-eclipse and out-of-eclipse segments with a two-component decay function. Such a joint treatment provides a more robust constraint on their relative scaling and allows the reprocessed fraction to be determined with lower uncertainty.}

\krrtext{In EXO 0748-676, the accretion disc is a plausible reprocessing site; however, even a maximally flared disc in EXO 0748-676 is expected to achieve only $\sim 2\%$ efficiency \citep{AHKnight2025_02}, assuming that all flux incident on the disc is re-emitted. In reality, part of the burst energy is absorbed and lost to heating, implying an even lower disc efficiency. Thus, the disc alone cannot account for the observed fractions. \cite{AHKnight2025_02} have considered another reprocessing agent, ablated material from the surface of the companion star. Observations of EXO 0748-676 report average average column densities of $\sim 10^{21}$-$10^{22}~\mathrm{cm^{-2}}$ \citep{Neilsen2023, AKnight2022}, but values up to $\sim 8\times 10^{24}~\mathrm{cm^{-2}}$ occur near eclipse transitions, within $\sim 700$ - $1500~\mathrm{km}$ of the companion star \citep{AKnight2022}. This highly ionised, dense region - likely produced by irradiation-driven ablation from the companion - could scatter burst radiation into the line of sight with higher efficiency. While such dense ablated material provides a plausible explanation for the higher reprocessing fractions observed, quantifying its contribution remains challenging due to uncertainties in the geometry, extent, and density structure of the outflow. Moreover, the ablated material may evolve dynamically on short timescales. Reprocessing by a disc wind could also contribute to the observed reflection. The detection of Fe~K$\alpha$ absorption lines supports the presence of a magnetohydrodynamic (MHD) disc wind \citep{Ponti2014, AHKnight2025_02}. However, significant uncertainties in key wind parameters - such as the mass-loss rate, geometry, and launch radius - make it difficult to accurately estimate its contribution to the reprocessing fraction. Furthermore, column densities measured away from eclipse transitions remain relatively low ($\sim 10^{21}$-$10^{22}~\mathrm{cm^{-2}}$), arguing against a persistent, high-density disc wind in EXO~0748-676. The only eclipse burst detected in XTE~J1710-281 shows a reprocessing fraction of $\sim$2.7\%. Given its similar orbital properties to EXO~0748-676, even a maximally flared disc alone is unlikely to account for the observed efficiency once absorption and re-radiation within the disc are considered. Observations of XTE~J1710-281 report an average column density of $\sim$$2\times10^{21}$~cm$^{-2}$ \citep{Neilsen2023}; however, disc winds with column densities up to $\sim$$7\times10^{23}$~cm$^{-2}$ have also been reported \citep{Younes2009, Raman2018}. Such high-density winds could scatter burst radiation into the line of sight with greater efficiency. A combination of a high-column-density disc wind and the accretion disc could plausibly explain the observed reprocessing fraction; however, variations in column density between deep and shallow pre-eclipse dips \citep{Younes2009, Raman2018} indicate that the disc wind is variable, making its contribution difficult to quantify. Further observations of eclipse-transition bursts, along with high-resolution studies of the dense winds in these system, are essential for constraining these uncertainties and refining reprocessing models.}

\krrtext{In EXO 0748-676, the reprocessing fraction shows an increasing trend for bursts occurring closer to eclipse egress (see Figure~\ref{fig:reprocessing_eff_02}). For reprocessing from the high-density ablated wind of the companion star, only the portion of the wind visible along the line of sight contributes. As the neutron star approaches eclipse egress, a larger fraction of this visible wind would become illuminated, thereby increasing the reprocessing fraction. Similarly, in the case of disc reprocessing, a larger portion of the accretion disc would become visible as the neutron star moves closer to egress, potentially enhancing the reprocessing fraction. Thus, the observed increase in reprocessing fraction for bursts occurring closer to eclipse egress is consistent with reprocessing from both the ablated wind of the companion star and the accretion disc.}

\krrtext{The fastest variability in the X-ray light curve sets an upper limit on the size of the reprocessing region, since variability on timescales shorter than the light-travel time across the region cannot be observed. In EXO 0748–676, the light-travel time across the system is $\sim$3 s \citep{hynes2006}, and burst rise times as short as 5-6 s are observed, giving an upper limit on the size of the reprocessing region that is larger but comparable to the size of the binary system itself. This remains consistent with reprocessing from the ablated wind near the companion star and the accretion disc.
For XTE J1710–281, with a binary orbital period of 3.28 h \citep{Jain2011}, and assuming a companion mass of 0.22 $M_\odot$ and neutron star mass of 1.4 $M_\odot$ \citep{Iaria2024}, Kepler’s third law gives a binary size of $\sim$3 light-seconds. The burst rise time of the only eclipse burst observed in this source is $\sim$4 s, implying an upper limit on the reprocessing region comparable to the binary size, again consistent with reprocessing in the accretion disc and disc wind.}

\section{\krrtext{Summary and conclusion}}
\label{eclipse_burst_conclusion}

\krrtext{We have investigated reprocessed thermonuclear bursts detected during the eclipses of the LMXBs EXO 0748-676 and XTE J1710-281. We estimate the reprocessing fractions in these systems at various orbital phases by modeling the peculiar eclipse bursts detected in these systems, with tails extending beyond the eclipses.}

\krrtext{
\begin{itemize}
    \item In EXO 0748–676 and XTE J1710–281, reprocessing from the accretion disc alone is unlikely to be sufficient to produce the observed reprocessing fractions. In EXO 0748–676, the ablated wind near the companion star together with the accretion disc are plausible contributors, while in XTE J1710–281 the accretion disc and disc wind are possible contributors. The increasing reprocessing fraction in EXO 0748–676 for bursts closer to eclipse egress is consistent with reprocessing from both the ablated wind and the accretion disc.
    \item Eclipse bursts in EXO 0748–676 require two exponential decay components to obtain good fits. The presence of two decay components could indicate a complex composition of the accreted fuel.
    \item The fastest burst rise times observed in EXO 0748–676 and XTE J1710–281 are comparable to the light-travel time across the binary system, which is consistent with reprocessing from the accretion disc, the companion star, and high-density winds within the system.
\end{itemize}
}

%%%%%%%%%%%%%%%%%%%%%%%%%%%%%%%%%%%%%%%%%%%%%%%%%%%%%%%%%%
\section*{Acknowledgements}
 \krrtext{We thank the anonymous referee for their valuable and constructive inputs, which have contributed to the improvement of the paper.} This research has made use of archival data and software provided by NASA's High Energy (HEASARC), which is a service of the Astrophysics Science Division at NASA/GSFC. VJ acknowledges the support provided by the Department of Science and Technology (DST) under the “Fund for Improvement of S \& T Infrastructure(FIST)” program (SR/FST/PS-I/2022/208). VJ also thanks the Inter-University Centre for Astronomy and Astrophysics (IUCAA), Pune, India, for the Visiting Associateship.
%Astrophysics Science Archive Research Center (HEASARC)\footnote{\url{https://heasarc.gsfc.nasa.gov/}}, which is a service of the Astrophysics Science Division at NASA/GSFC. 

%%%%%%%%%%%%%%%%%%%%%%%%%%%%%%%%%%%%%%%%%%%%%%%%%%%%%%%%%
\section*{Data Availability}
The observational data underlying this work is publicly available through the High Energy Astrophysics Science Archive Research Center (HEASARC). Any additional information will be shared on reasonable request to the corresponding author.



\bibliographystyle{mnras}
\bibliography{reference_lmxb} 

%%%%%%%%%%%%%%%%%%%%%%%%%%%%%%%%%%%%%%%%%%%%%%%%%%
\appendix
\section{Appendix}

\begin{figure}
  \centering
  \includegraphics[width=8cm]{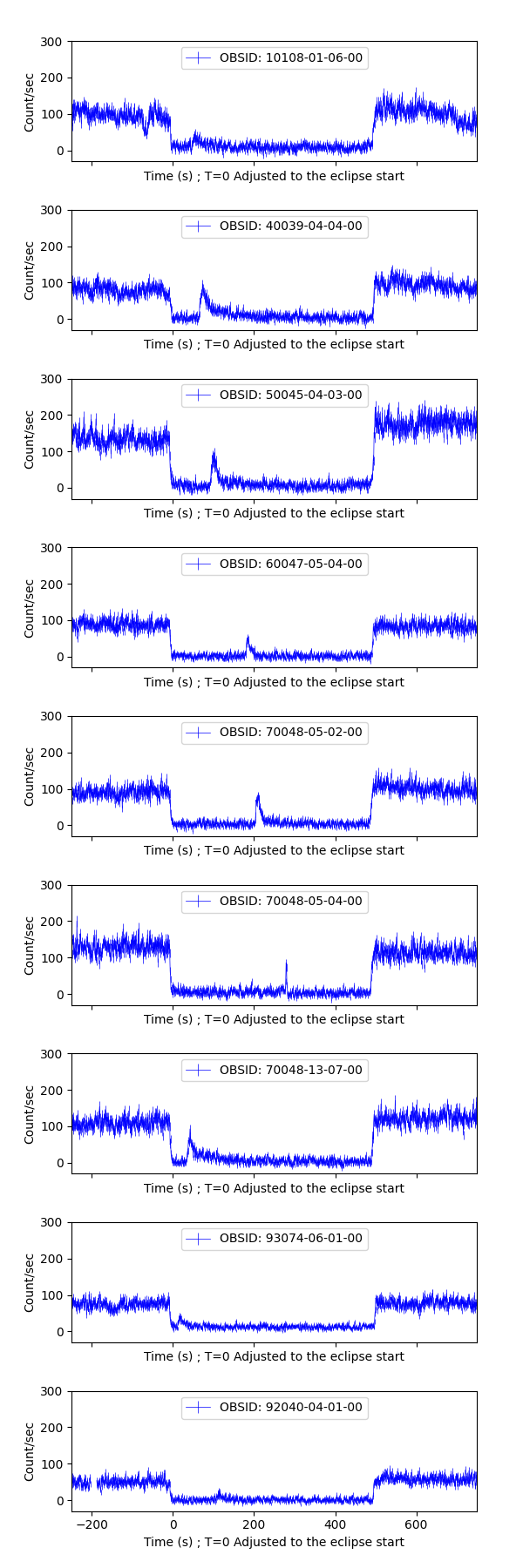}
  \caption{Eclipse bursts detected in \textit{RXTE} data of EXO 0748-676 that occurred entirely during the eclipse}
  \label{fig:exo_0748_676_eclipse_plots}
\end{figure}

\begin{figure}
  \centering
  \includegraphics[width=8cm]{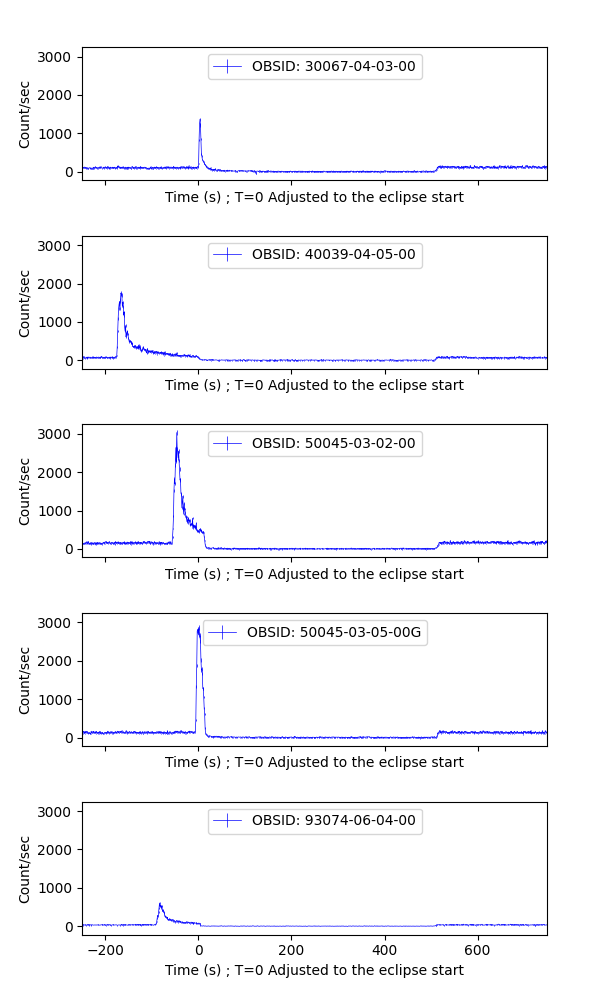}
  \caption{Bursts detected in \textit{RXTE} data of EXO 0748-676 with their tails extending into the eclipse duration}
  \label{fig:exo_0748_676_eclipse_ingress_plots}
\end{figure}

\begin{comment}
\begin{figure}
  \centering
  \includegraphics[width=8cm]{exo_0748_676/exo_0748_676_eclipse_egress_plots.png}
  \caption{Eclipse bursts detected in \textit{RXTE} data of EXO 0748-676 with their tails extending beyond the eclipse duration}
  \label{fig:exo_0748_676_eclipse_egress_plots}
\end{figure}
\end{comment}

\begin{figure}
  \centering
  \includegraphics[width=8cm]{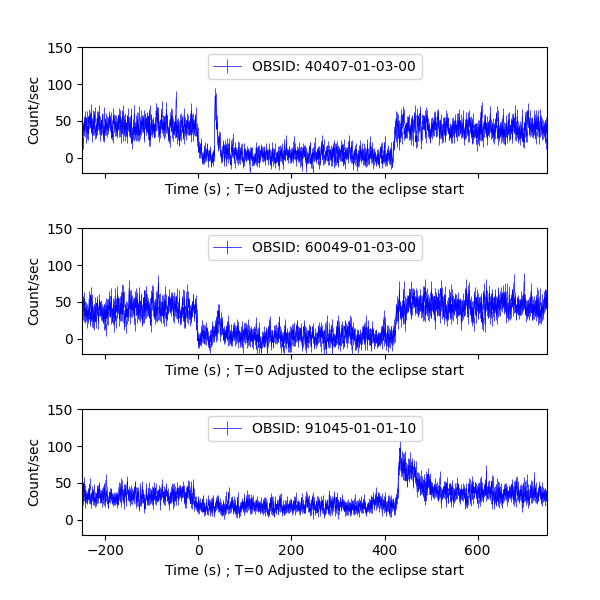}
  \caption{Eclipse bursts detected in \textit{RXTE} data of XTE J1710-281}
  \label{fig:xte_j1710_281_eclipse_plots}
\end{figure}

%%%%%%%%%%%%%%%%%%%%%%%%%%%%%%%%%%%%%%%%%%%%%%%%%%

% Don't change these lines
\bsp	% typesetting comment
\label{lastpage}
\end{document}